\documentclass[aip,reprint]{revtex4-1}
\usepackage[version=3]{mhchem} 
\usepackage{natbib}
\usepackage{soul}
\usepackage{color}
\usepackage{epstopdf}
\usepackage{graphicx}

\begin{document}
\title{Deposition of thin silicon layers on transferred large area  CVD graphene}

\author{Grzegorz Lupina}
\email{lupina@ihp-microelectronics.com}
\author{Julia Kitzmann}
\author{Mindaugas Lukosius}
\author{Jarek Dabrowski}

\author{Andre Wolff}
\author{Wolfgang Mehr}
\affiliation{IHP, Im Technologiepark 25, 15236 Frankfurt (Oder), Germany}

\begin{abstract}

Physical vapor deposition of Si onto transferred CVD graphene is investigated. At elevated temperatures Si nucleates preferably on wrinkles and multilayer graphene islands. In some cases, however, Si can be quasi-selectively grown only on the monolayer graphene regions while the multilayer islands remain uncovered. Experimental insights and ab initio calculations show that variations in the removal efficiency of carbon residuals after the transfer process can be responsible for this behavior. Low-temperature Si seed layer results in improved wetting and enables homogeneous growth. This is an important step towards realization of electronic devices in which graphene is embedded between two Si layers.

\end{abstract}

\maketitle

Graphene-Si junctions\cite{tongayPRX2012} gain increasing attention for applications in infrared photodetectors\cite{Lv2013}, solar cells\cite{solarcellsMiao2013}, and gate controlled variable Schottky barrier transistors\cite{barristor2012}. Prototypes of such devices are so far build by transferring graphene onto crystalline Si substrates. Although transfer of graphene may be a viable option for some applications, it is not a generally preferred solution in microelectronic manufacturing where a direct deposition method would be ideal \cite{reviewferrari2012,Lippert2013Carbon}. Once available such a method will significantly accelerate the integration of graphene into the mainstream Si technology, however, fabrication of high-quality graphene-Si interfaces will most probably remain challenging due to high reactivity between C and Si and SiC formation\cite{CSiPhasediag}. The opposite scenario, in which Si is deposited onto graphene to form Si-graphene junction can thus be more attractive and inevitable to realize promising terahertz electronic device concepts in which graphene is embedded between two Si layers such as in the graphene-base heterojunction transistor\cite{dilecce2013}. As of this writing, the literature reports on the attempts to grow Si on graphene are scarce. 

Here, using physical vapor deposition we explore the growth of Si layers on Si/SiO$_2$ substrates covered with transferred chemical vapor deposited (CVD) graphene. We find that Si films grown directly at elevated temperatures are clearly discontinuous and that application of a low-temperature seed layer improves wetting so that closed Si films can be obtained. We show that there are significant differences in wettability of monolayer regions and multilayer graphene islands which can be due to differences in the efficiency of polymer residuals removal on these areas. Results presented in this work contribute to the understanding and solution of problems associated with cleaning of CVD graphene and the deposition of various materials on graphene.

Commercially available graphene was transferred from Cu using a standard poly(methyl methacrylate) (PMMA)-assisted method\cite{liang2011,TransferRuoff2011} onto patterned Si(100) substrates with Si pillars embedded in SiO$_2$. After transfer, PMMA layer was removed in acetone and the sample was loaded into an ultra-high vacuum (UHV) molecular beam epitaxy system. Before Si deposition samples were annealed in UHV at $550^\circ$C for 10 min. Growth was performed at a pressure of $2\times$ $10^{-8}$ mbar from a high purity source heated by electron beam. 
\begin{figure}[h]
\includegraphics[width=1\columnwidth]{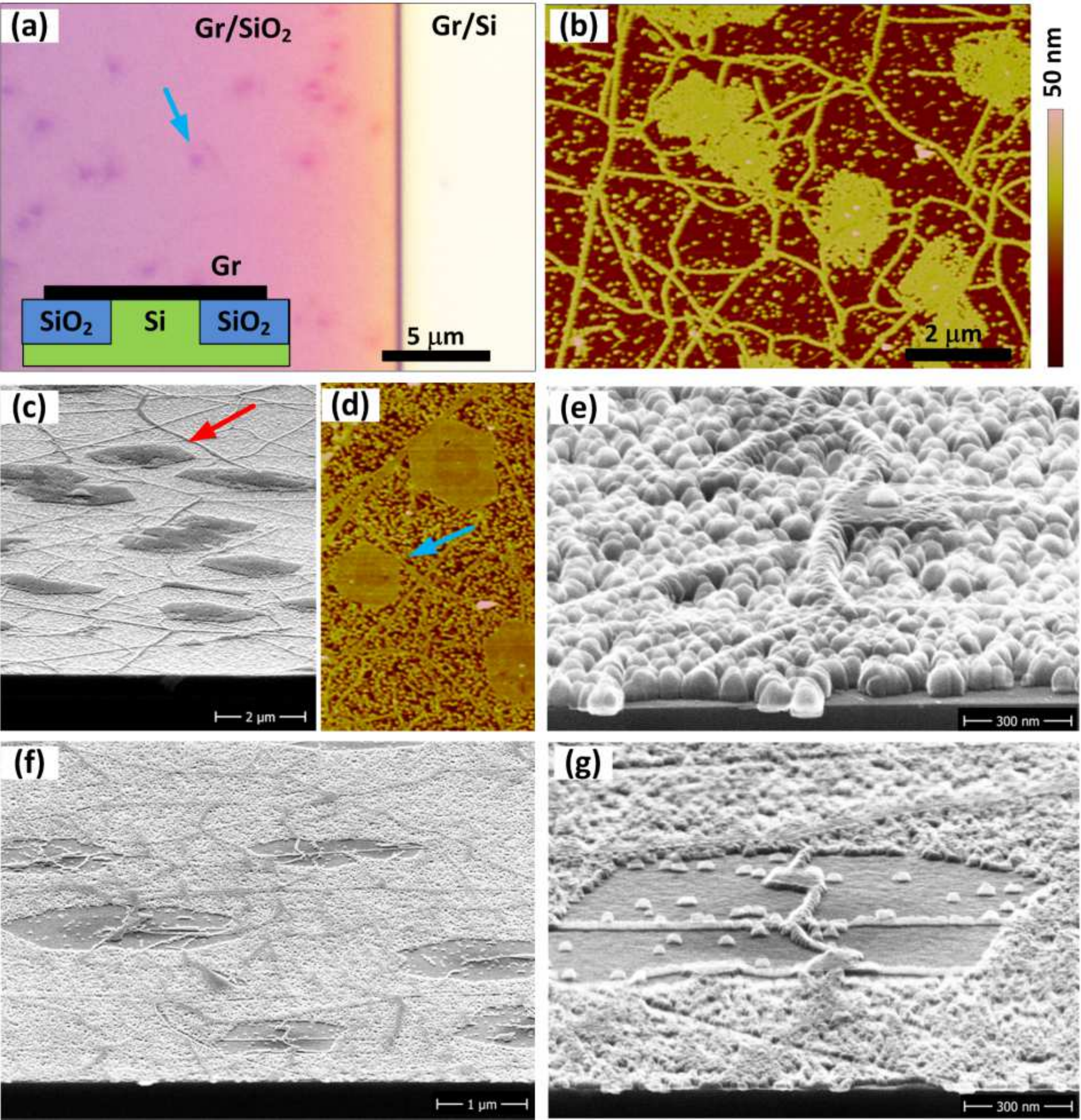}

\vspace{-10pt}
\caption{\label{fig1} (a) Optical microscope image of graphene layer transferred to SiO$_2$/Si substrate. Arrow indicates a multilayer graphene island. The inset shows a schematic cross-section of the substrate. (b) AFM image after deposition of 25 nm of Si. (c-d) SEM and AFM image after Si deposition showing hexagonally shaped islands covered with Si. (e) SEM image of a 100 nm thick Si layer in the monolayer graphene region. (f-g) SEM images after Si deposition on a different sample. Hexagonally-shaped regions not covered with Si can be distinguished.}
\end{figure}
Optical microscope image of a graphene layer transferred on patterned substrate is shown in Figure 1(a). The inset shows a schematic cross-section of the substrate. The graphene layer along with dark multilayer islands\cite{bilayerRuoff2012,bilayerislandsLiu2012} can be distinguished on the areas covered with SiO$_2$ but it is not visible on the Si pillars. As we noted in our previous work\cite{Lupina2013}, the number, shape, and distribution of the multilayer islands on graphene depends very much on the graphene supplier and can be easily evaluated on SiO$_2$/Si substrates using optical microscope. On SiO$_2$/Si substrates Raman signals of graphene are enhanced with respect to clean Si substrates\cite{RamEnhYoon2009} and for this reason Raman measurements in this work were performed mainly on the SiO$_2$ covered areas (SiO$_2$ thickness of about 470 nm). Atomic force microscopy (AFM) and secondary electron microscopy (SEM) images were acquired on both Si and SiO$_2$ parts showing no qualitative difference in the Si growth behavior.
\newline 
AFM image taken after deposition of nominally 25 nm Si at the substrate temperature of $550^\circ$C is shown in Figure 1(b). The surface of the sample is very inhomogeneous. The majority of Si atoms is deposited on graphene wrinkles (bright irregular lines) and oval-shaped areas which are associated with multilayer graphene islands. On the islands, growing Si layer is relatively flat with rms roughness of 1-3 nm. Between the elevated islands rms roughness is much larger (8-10 nm). This result is similar to the one obtained in our recent experiments with CVD grown HfO$_2$ on graphene\cite{Lupina2013}. The nucleation pattern on monolayer graphene regions resembles formation of Si nanoclusters on highly oriented pyrolytic graphite studied previously with AFM and scanning tunneling microscopy \cite{Scheier2000,aSiIkuta1994,Buuren1998,aSi_Matsuse_1996}. Depending on the choice of graphene material, the multilayer islands preferentially covered with Si are either oval-shaped (e.g. Fig. 1(b)) or regularly hexagonally-shaped as shown in Figures 1(c) and (d). The shape of the latter features bear a close resemblance to the individual graphene grains in the initial growth stage on Cu\cite{Yu}. Regardless of the shape of the islands, the Si layer nucleates much better on these areas than between them: on the monolayer regions growth seems to proceed by an island-like Volmer-Weber mode and even a nominally 100 nm thick Si layer is discontinuous as demonstrated by the SEM image in Fig. 1(e). Interestingly, on some samples an exactly opposite nucleation scenario is observed. This is illustrated in Figures 1(f) and (g). Here again 25 nm of Si is deposited at $550^\circ$C on graphene sample prepared in the same way as described above. However, in contrast to the case presented in Fig. 1(c), Si grows mostly on the monolayer graphene regions and wrinkles leaving thicker hexagonally-shaped graphene islands practically uncovered. Apparently in this case there are no or only very few nucleation sites available on the islands. 
It has been shown, that during atomic layer deposition (ALD) and CVD on graphene nucleation takes place preferentially at the edges and wrinkles i.e. at the places where defects providing pinning sites are expected to occur\cite{nucleationWang2008,Lupina2013}. To verify if a similar mechanism is involved in the peculiar growth behavior observed here we performed Raman spectroscopy measurements shown in Figure 2. Raman spectra presented in Figure 2(a) were acquired on the places indicated in Figures 2(b)-(d) showing microscope images of bare graphene substrate, sample with no Si deposit on the islands, and sample with homogeneous nucleation on the islands, respectively.

\begin{figure}[h]
\includegraphics[width=0.8\columnwidth]{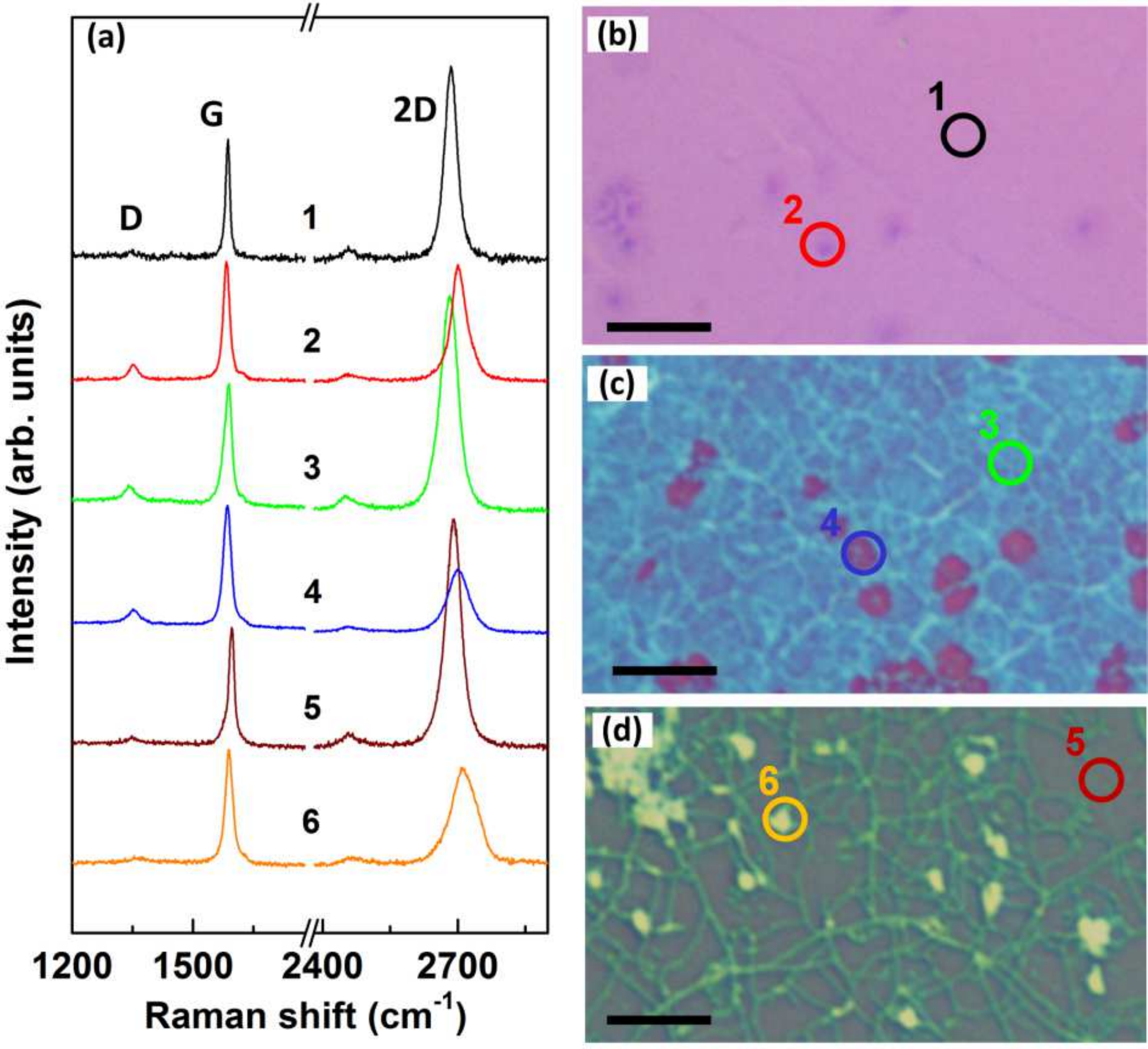}

\vspace{-10pt}

\caption{\label{fig2} (a) Raman spectra acquired from the points indicated in optical microscope images shown in panels b-d. Spectra are normalized to the same G mode intensity and vertically offset for clarity. (b) graphene after transfer. (c) after Si deposition with no Si growth on the multilayer islands. (d) after Si deposition resulting in uniform coverage of multilayer islands with Si. Scale bar in b-d is 5 $\mu$m.}
\end{figure}

Spectra labeled with odd numbers are acquired on the monolayer regions; those labeled with even numbers are collected from the multilayer islands. Measurements on the graphene islands systematically show lower 2D/G intensity ratios and broader 2D bands if compared to the scans on the monolayers. However, there is no clear correlation between the intensity of the defect-related D-band and the coverage of the islands with Si. In other words, a homogeneous growth of Si on multilayer islands does not clearly correlate with a significantly increased number of defects detectable by Raman on these sites.

\begin{figure}[h]
\includegraphics[width=0.8\columnwidth]{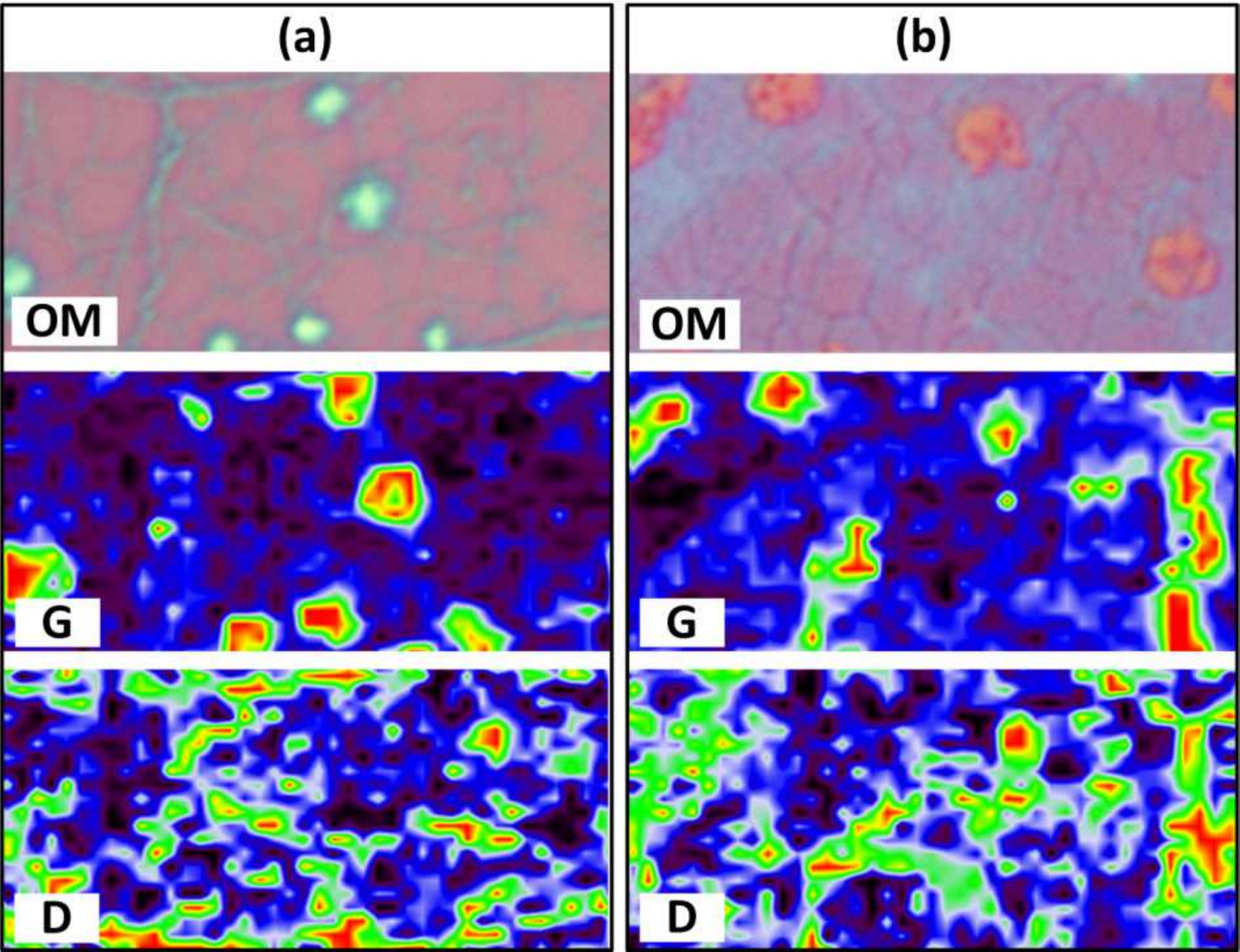}

\vspace{-10pt}

\caption{\label{fig3} Optical microscope images and Raman spectroscopy mapping on two samples with multilayer islands covered with Si (a) and multilayer graphene islands remaining uncovered after Si deposition (b). Analyzed area is 10$\times$20 $\mu$m$^2$. Objective with numerical aperture of 0.85 and 514 nm laser light was used.}
\end{figure}

To avoid non-representative spots and strengthen the above conclusion we performed Raman mapping with high spatial resolution on area of 10$\times$20 $\mu$m$^2$ of two samples showing different Si growth behaviors. These results are presented in Figure 3. There is a strong correlation between the position of the covered/uncovered islands and the intensity of the G (and 2D, not shown) mode: G peak is usually more intense on the islands regardless of the analyzed case. For the D-mode the correlation, if any, is much weaker. According to these results, the quasi-selective growth observed on some samples does not seem to be a consequence of different defect distributions and/or concentrations. At least not defects detectable by Raman spectroscopy\cite{Ferrari2013}.

\begin{figure}[h]
\includegraphics[width=0.9\columnwidth]{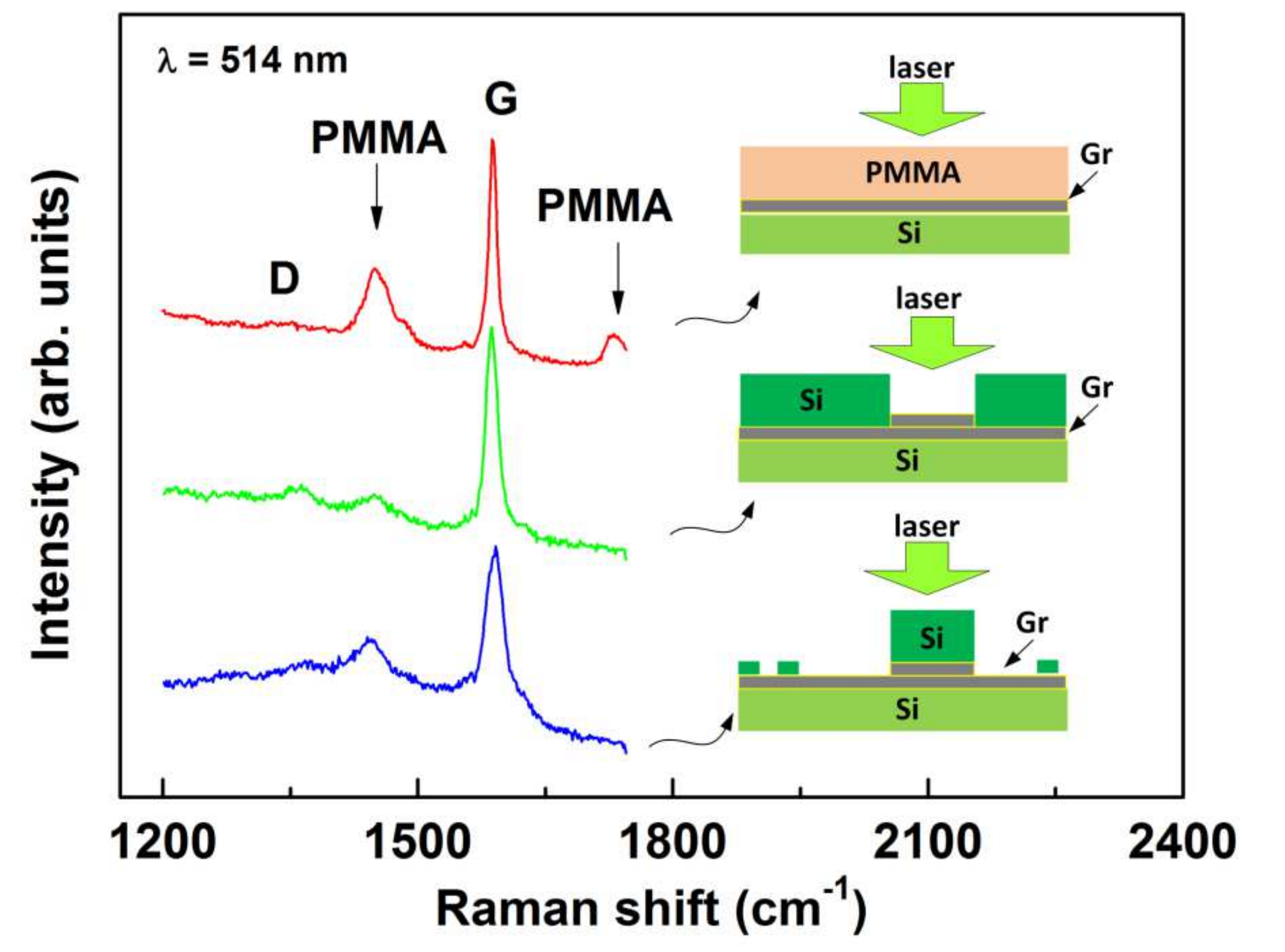}

\vspace{-10pt}

\caption{\label{fig4n} Raman spectra around the G mode acquired for a reference sample before PMMA removal in acetone compared with Raman spectra from two graphene samples covered with Si. For samples covered with Si measurements were performed on Si pillars with long integration times to reveal the presence of additional features possibly related to PMMA residues.}
\end{figure}

While the origin of the heterogeneous nucleation behavior remains under investigation, a plausible explanation can be a thin layer of transfer-related impurities which is in some cases not removed by acetone, isopropanol, and subsequent annealing in UHV. Presence of such a layer could functionalize the surface of graphene (islands) and improve nucleation of the Si layer. In fact, evidence supporting such a hypothesis is provided by Figure 4 showing additional Raman measurements performed on samples with poor (cf. Fig. 2(c)) and good (cf. 2(d)) nucleation on the multilayer islands. In contrast to the spectra presented in Fig. 2, here the measurements were conducted on graphene islands located above Si pillars. Since Raman signatures of graphene are relatively weaker on Si substrates\cite{RamEnhYoon2009} much longer integration times were required to obtain good signal/noise ratios. These measurements reveal presence of additional features in the spectra which can be assigned to PMMA residues and their derivatives. In particular, a PMMA-related mode due to CH$_2$ deformation \cite{PMMA_Raman_Polymer1969,Raman_PMMA_2011} is detected at 1450 cm$^{-1}$ for the reference sample before PMMA removal as well as for samples after PMMA removal, UHV cleaning, and Si deposition. On the islands where Si nucleates well (cf. Fig. 1(c)), additionally a broad Raman band is visible which can be due to residual carbons on the graphene layer\cite{CO2clean_Chabal_2013}. The islands characterized by poor Si nucleation (cf. Fig. 1(g)) appear to be cleaner as the spectral features related to the residues are weaker. In fact, the surface of these islands may resemble closely a clean graphite surface with a low sticking probability and large diffusion length of adsorbed Si atoms. In such a case, since there are no or very few nucleation sites available on the islands, Si atoms once adsorbed can travel long distances to the nearest wrinkle or single layer graphene region. There, suitable nucleation sites are available and the growth takes place. However, a limited spatial resolution of our experimental setup does not allow us to confirm clearly that the surface of multilayer graphene islands uncovered by Si is in fact completely free of PMMA residues.

\begin{figure}[h]
\includegraphics[width=0.9\columnwidth]{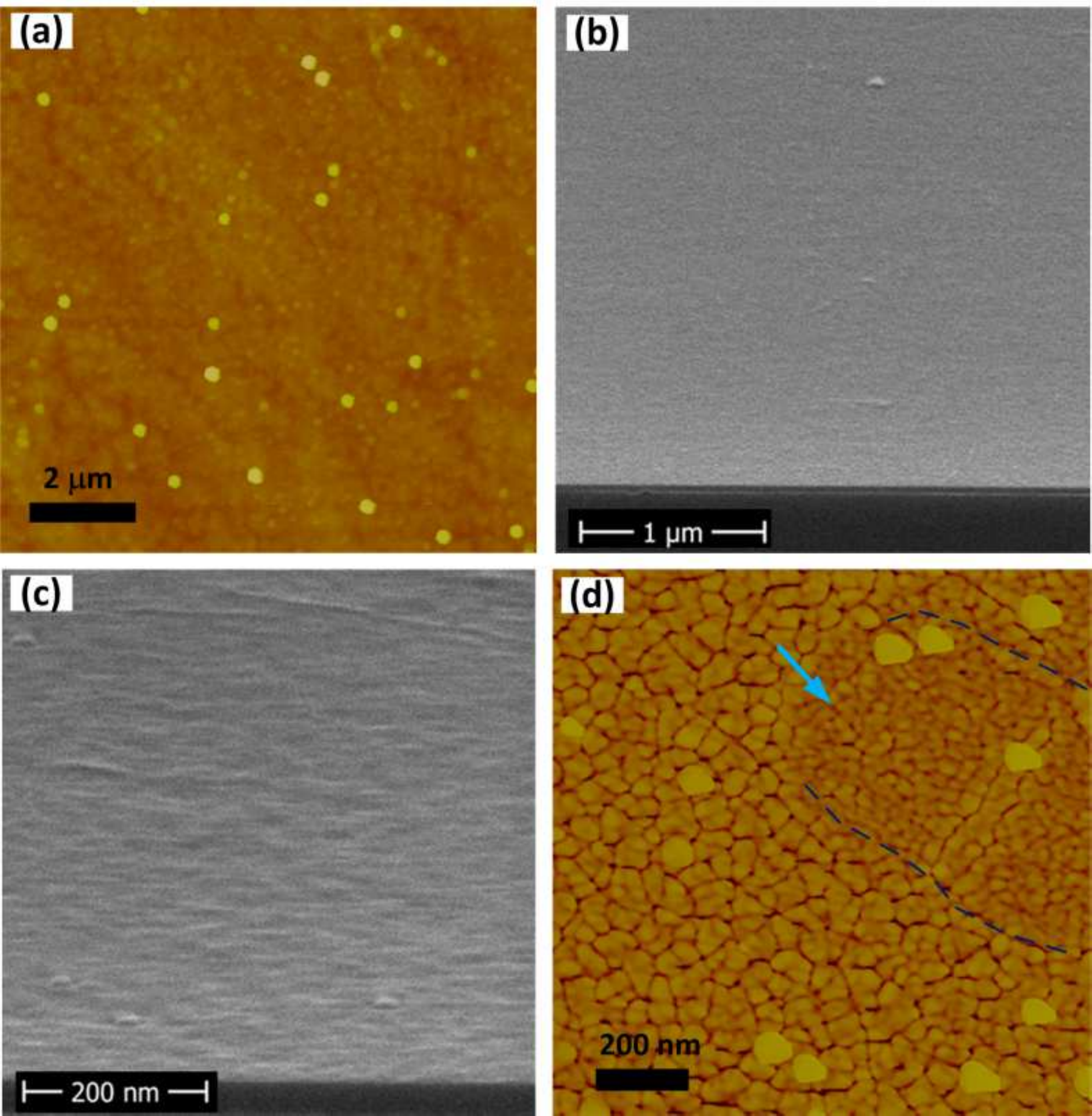}

\vspace{-10pt}

\caption{\label{fig4} (a) AFM image from a 15nm thick Si layer deposited on graphene at $50^\circ$C. (b-c) SEM image of a 25 nm thick Si layer deposited at $550^\circ$C on the low-temperature seed layer. (d) AFM phase image of the layer shown in panels b-c revealing regions with different surface topography.}
\end{figure}

This interpretation of the observations is corroborated by results of ab initio density functional theory calculations. Si atoms barely stick to perfect graphene. For Si on undoped graphene, the adsorption energy is 0.58\,eV plus Van der Waals contribution (expected to be of the order of 50\,meV), and the energy barrier for surface diffusion is 60\,meV, meaning that at $550^\circ$C the Si atom desorbs after only a few diffusion step on the surface. 
The adsorption energy changes to about 0.94\,eV (0.48\,eV) when graphene is p-type (n-type). Nevertheless, if the deposition is done at $550^\circ$C, only those atoms that fall about 20 to 200 nanometers away from a seed site or from the silicon island that already exists, can contribute to growth. This explains the growth mode at $550^\circ$C.

The calculated energy barriers indicate that lowering the growth temperature to $50^\circ$C would suffice to change the growth mechanism to that in which the atoms falling on perfect graphene spend enough time in the adsorbed state to diffuse to the growing island. Si deposited at $50^\circ$C on p-type graphene may even be able to nucleate on perfect regions of the substrate, unless small Si$_n$ clusters ($n=2,3,$\dots) are insufficiently bonded to the sheet.
To obtain a closed layer of Si on graphene we have therefore developed a two-step process in which a thin Si seed layer is firstly deposited at $50^\circ$C. 
In the second step, Si growth is continued at elevated temperatures. This is similar to the approach using a thin metal layer to facilitate ALD growth of dielectrics on graphene\cite{seedmetalKim2009,Sam2012Nanolett}. AFM image in Fig. 5(a) shows the surface of a 15 nm-thick seed layer. Beside occasionally occurring spots the surface is relatively uniform and flat with rms roughness of 0.4-0.5 nm. Moreover, at this substrate temperature there is no visible difference between the growth on multilayer islands and the monolayer regions. This seed layer significantly improves wetting during further deposition at $550^\circ$C. Figure 5(b) and (c) show SEM images of a 25 nm Si layer deposited at $550^\circ$C on the low-temperature seed. The layer is continuous and relatively homogeneous. Some inhomogeneity in surface topography is, however, observed in AFM images. Particularly, in the phase images oval-shaped areas with the size corresponding to the size of multilayer islands can be easily recognized as indicated in Fig. 5(d). The Si layer in these areas is smoother (0.3-0.4 nm rms) than outside these regions (0.6-0.9 nm rms). Apparently, the uniform seed layer undergoes reorganization during the high temperature step which results in small differences in the growth on multilayer islands and on the monolayer regions. Despite that the Si layer deposited at $550^\circ$C is closed and free of pinholes.

\begin{figure}[h]
\includegraphics[width=0.7\columnwidth]{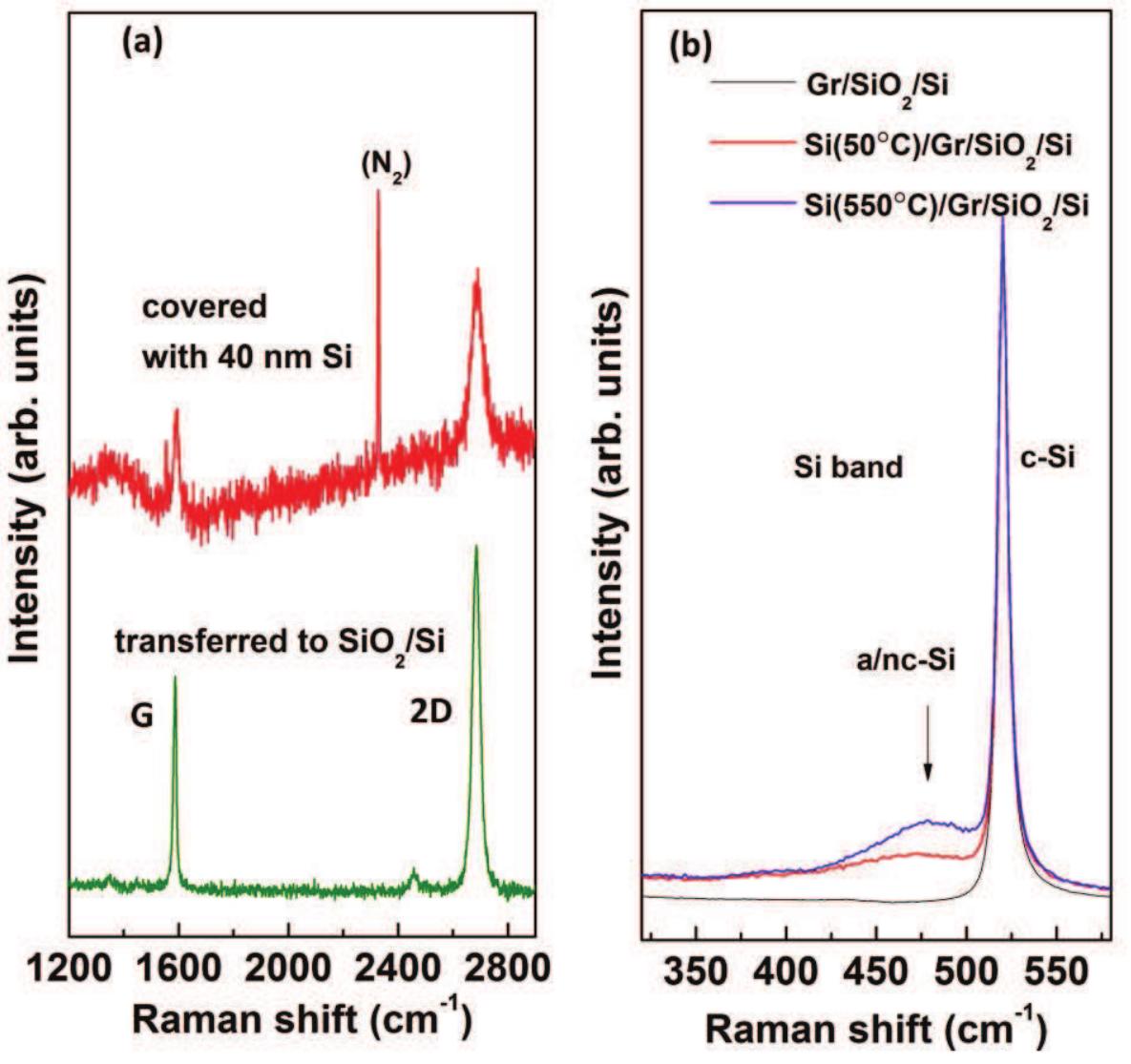}

\vspace{-10pt}

\caption{\label{fig5} (a) Raman spectra from graphene before and after Si deposition. The peak at 2331 cm$^{-1}$ is due to atmospheric N$_{2}$ which becomes visible due to long integration times. (b) Comparison of Raman Si-mode for various samples.}
\end{figure}

Figure 6(a) compares Raman spectra of the graphene layer before and after Si deposition. Although Raman signals after deposition are attenuated by the about 40nm thick Si layer, it can be seen that graphene largely preserves its quality throughout the Si deposition process. Scans performed around the position of the Si mode (Fig. 6(b)) indicate that the Si layer deposited in the two-step process is of nanocrystalline/amorphous nature\cite{Gaisler2013}.

In summary, physical vapor deposition of Si on transferred CVD graphene was investigated. The presence of multilayer graphene islands and wrinkles results in a heterogeneous nucleation scheme. Si nucleates poorly on the monolayer graphene regions and the growth seems to proceed by an island-like Volmer-Weber mode while on the multilayer islands growth is more two-dimensional with uniform coverage. Interestingly, in some cases the growth on multilayer islands is not observed at all resulting in a quasi-selective deposition only on the monolayer regions. According to the results presented here, it can be due to a more effective removal of polymer residues from the surface of multilayer islands. Further studies are under way to clarify the exact origin of this behavior which can potentially lead to a solution of problems associated with deposition of various materials on graphene. Meanwhile we have shown that a low-temperature seed layer is effective in improving Si growth at elevated temperatures and enables closed and smooth Si layers on graphene. This opens the way to the realization of novel electronic devices consisting of graphene embedded between two Si layers.

\vspace{10pt}

The authors thank G. Morgenstern, H.-P. Stoll, H.-J. Thieme, G. Lippert, M. Fraschke, and Y. Yamamoto for experimental support and discussions. Financial support by the German Research Foundation (DFG, Project No. ME 4117/1-1) and the European Commission through a STREP project GRADE (No. 317839) is gratefully acknowledged. We gratefully acknowledge the computing time granted by the John von Neumann Institute for Computing (NIC) and provided on the supercomputer JUROPA at Julich Supercomputing Centre (JSC) to the project hfo06.

\end{document}